# Folding of guanine quadruplex molecules – funnel-like mechanism or kinetic partitioning? An overview from MD simulation studies.

Jiří Šponer,[a,b]* Giovanni Bussi,[c] Petr Stadlbauer,[a,b] Petra Kührová,[b] Pavel Banáš,[b] Barira Islam,[a] Shozeb Haider,[d] Stephen Neidle,[d] Michal Otyepka[b]

[a] Institute of Biophysics, Academy of Sciences of the Czech Republic, Kralovopolska 135, 612 65 Brno, Czech Republic
[b] Regional Centre of Advanced Technologies and Materials, Department of Physical Chemistry, Faculty of Science, Palacky University Olomouc, 17. listopadu 12, 771 46 Olomouc, Czech Republic
[c] Scuola Internazionale Superiore di Studi Avanzati, Via Bonomea 265, 34136 Trieste, Italy
[d] UCL School of Pharmacy, 29-39 Brunswick Square, London WC1N 1AX, UK

* Corresponding author, sponer@ncbr.muni.cz

**Abstract**

**Background**

Guanine quadruplexes (GQs) play vital roles in many cellular processes and are of much interest as drug targets. In contrast to the availability of many structural studies, there is still limited knowledge on GQ folding.

**Scope of review**

We review recent molecular dynamics (MD) simulation studies of the folding of GQs, with an emphasis paid to the human telomeric DNA GQ. We explain the basic principles and limitations of all types of MD methods used to study unfolding and folding in a way accessible to non-specialists. We discuss the potential role of G-hairpin, G-triplex and alternative GQ intermediates in the folding process. We argue that, in general, folding of GQs is fundamentally different from funneled folding of small fast-folding proteins, and can be best described by a kinetic partitioning (KP) mechanism. KP is a competition between at least two (but often many) well-separated and structurally different conformational ensembles.

**Major conclusions**

The KP mechanism is the only plausible way to explain experiments reporting long time-scales of GQ folding and the existence of long-lived sub-states. A significant part of the natural partitioning of the free energy landscape of GQs comes from the ability of the GQ-forming sequences to populate a large number of *syn-anti* patterns in their G-tracts. The extreme complexity of the KP of GQs typically prevents an appropriate description of the folding landscape using just a few order parameters or collective variables.

**General significance**

We reconcile available computational and experimental studies of GQ folding and formulate basic principles characterizing GQ folding landscapes.

## 1. Introduction

Correct folding of biomolecules is a crucial step in many biochemical processes [1, 2]. However, there are no experimental methods to monitor biomolecular folding at the atomistic level of resolution. Nevertheless, for fast-folding short proteins that fold on a sub-millisecond time-scale by means of a funnel mechanism [3], unbiased (see section 2.1) atomistic explicit solvent molecular dynamics (MD) simulations have visualized structural details of folding of the individual molecules [4-8]. Although not fully converged, and although often performed at the melting temperature, the simulations provided a sufficient number of individual single-molecule folding and unfolding events that allow atomistic understanding of the process.

For nucleic acids, unbiased atomistic folding simulations of the shortest DNA hairpins with Watson-Crick stems suggested fast µs-scale folding slowed-down by longer-lived off-pathway intermediates [9]. Folding of short RNA hairpin tetraloops (TLs) using enhanced-sampling temperature replica exchange MD simulations (T-REMD, see section 2.5.1) have also been reported [10], though subsequent studies by other groups opened some issues on the convergence of T-REMD simulations for these system [11, 12]. Further, these works have shown that none of the presently available RNA force fields is fully satisfactory in order to simulate RNA TL folding [11, 12]. Additionally, it is difficult to extract unbiased kinetic information from T-REMD simulations.

Compared to the above-noted systems, folding of guanine quadruplexes (GQs) is much more complicated. Below, we summarize what has been learnt about GQ folding and unfolding pathways using MD simulations. We first provide a conceptual overview of the picture of GQ folding that is quite consistently emerging from computations. Then we explain specific results together with a short description of the methods, their limitations and future perspectives. Throughout the paper, we discuss the computed results in the context of the available experiments.

### 1.1 Funnel vs. kinetic partitioning.

The term folding funnel has sometimes been also used in connection with the folding of intramolecular DNA GQs. However, we suggest that the word 'funnel' is misleading for GQs. A funneled free-energy landscape is characterized by a smooth decrease of the energy and configurational entropy, with the molecules decisively gliding towards the native state with continuous increases in the number of native contacts as the transition progresses [1, 4-8]. Along the pathway, there are no kinetic traps with local energy minima significantly deeper than the energy of thermal fluctuations. The funneled folding landscape is only marginally frustrated by non-native interactions and leads to fast folding events. This is not the case of GQ DNA sequences, where many experiments have revealed folding on a time-scale of up to several days [13-23].

Based on polymer physics, the speed of funnel-like folding should depend on the number of residues and would thus be extremely fast for GQs having ~25 nucleotides [1]. However, random heteropolymers do not fold via a funnel. The funnel landscapes of small fast-folding proteins result from either evolution or targeted design. Proteins have twenty possible residue types (amino acids) at each chain position to optimize the folding. In contrast, GQs have invariably only Gs in their G-tracts, and the remaining residues are limited to four types only. Furthermore, protein folding and stability are driven by local native contacts, while the topologically dominant native GQ interactions are the G-quartets, i.e., non-local interactions (assisted by ions to get the optimal free energy). This suggests that the GQ folding is likely different from the folding of small proteins. As we will argue below [24-27], and as suggested by others [13, 14, 22, 23], the time-scale of GQ folding in those experiments reporting long-time-scale folding [13-16, 22, 23] and long-lived substates [17, 22] can only be explained by a kinetic partitioning (KP) mechanism, i.e., a multi-pathway process over a rugged free-energy landscape [1, 28].

The KP folding landscape is the opposite of the funneled landscape. It contains deep competing free-energy minima (alternative folds, competing conformational basins or ensembles) separated by large free-energy barriers. Only a fraction of molecules folds directly to the native basin (native basin of attraction, NBA), i.e., the one which is the most populated at the thermodynamic equilibrium. The other molecules in the ensemble initially fold into some competing (non-native) basins (competing basin of attraction, CBA). The molecules become trapped at different basins and thermodynamic equilibrium is only reached after a sufficient number of misfolding-unfolding single-molecule events that ultimately lead to the equilibrium population of all the basins (Figure 1). The CBAs make the process slow. In principle, a KP mechanism may involve only two basins (the NBA and one CBA). However, in practice such a situation is unlikely. More common is that the folding landscape consists of several (or even numerous) well-separated deep basins and even the final folded state may consist of more than one basin with detectable population. For human telomeric GQ DNA sequences multiple different folds can co-exist at the thermodynamic equilibrium [13, 18, 19, 29-46]. Even when the native state in the equilibrium involves only one detectable basin, other basins can be transiently populated during the folding process. Additionally, relative stabilities of different basins may be dramatically influenced by the environment.

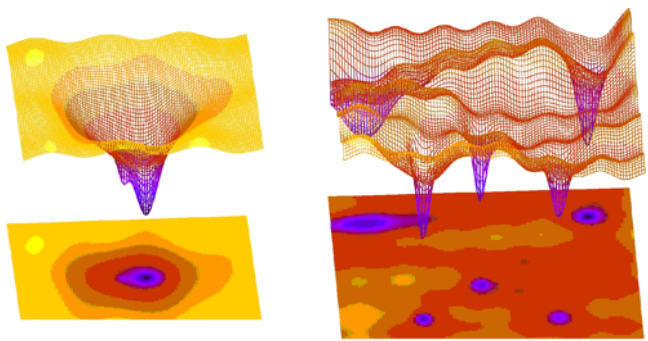

**Figure 1**. Funnel landscape (left) vs. kinetic partitioning (KP, right). Funnel landscape leads to rapid folding into only one free-energy basin on the folding landscape from any unfolded ensemble, i.e., there are no competing off-pathway intermediates. In addition, a true funnel folding is not significantly slowed down by on-pathway non-native interactions. KP landscape consists of multiple competing basins separated by large free-energy barriers. They may have different kinetic accessibilities from different unfolded ensembles. The populations of different competing basins typically vary with time as the process progresses towards the thermodynamic equilibrium. Multiple basins can be significantly populated at the equilibrium and form the native state, many others can be present only temporarily during the process.

Experimental identification of multiple competing folds populated during the folding process but vanishing at the thermodynamic equilibrium is difficult, since they may mutually overlap in the measurable signals during the folding, making them unresolvable. We would even suggest that there are essentially no experimental methods that would allow reliable monitoring of many CBAs during the folding process, so the true complexity of the folding processes remains hidden. It is also notable that knotty KP folding may be inherently not describable by simple concepts such as transition paths and transition state ensembles [47-49]. As will be discussed below, such processes are deemed to be not reducible to a few order parameters or (in computations) a few collective variables, despite attempts to use such simplified descriptions to describe at least portions of the landscape.

KP folding processes are considerably more challenging to investigate than fast funnel-like folding. The principles of movements among different basins are not fully understood at the atomistic level of description. It is not clear to what extent the misfolded molecules need to unfold before being able to move to another basin. It is important to point out that the distinction between folded, misfolded and unfolded states (structures) depends on the resolution of particular methods and on their definitions. In the real molecular world, individual molecules are sampling a continuum of geometries within their available conformational (phase) space, obeying the Boltzmann distribution at equilibrium. The main basins are those regions of conformational space where the molecules accumulate (i.e., reside with high probability) relative to the surrounding space. The inter-basin transitions likely involve a rich spectrum of structural dynamics. They may combine sudden fast rearrangements (barrier-crossing events) through some (pseudo)transition states between pairs of basins with slow conformational diffusion through the conformational space. In the latter case the molecules would travel from one sub-state to another through a series of smaller rearrangements with step by step (and back and forth) restructuring of H-bond networks and other molecular interactions. As already noted, KP (multiple-pathway) folding mechanisms are naturally unsuitable for description using simplified concepts of reaction coordinates or transition state ensembles, as the process may be principally non-projectable to a limited set of collective variables (see section 2.5.2) [47].

While we do not rule out that under specific experimental conditions a GQ may avoid the KP mechanism of folding, it would not be trivial for this to occur. The fact that many GQs fold slowly and that many experiments detect long-lived states [13-17] means that the KP mechanism is robustly inherent to the conformational space (folding landscape) of many GQ DNA sequences. Typical signs of partitioning are the equilibrium coexistence of different structures or occurrence of diverse native folds for similar sequences in different experimental conditions (as common for the human telomeric DNA sequence). The polymorphism does not need to be detectable in equilibrium, but it may be still present during the folding, leading to slow kinetics. Smooth fast folding of all the DNA GQ molecules in the ensemble directly to a single conformation would thus require specific restraints on the unfolded state from which the folding process is initiated. If a specific experiment indicates very fast GQ folding, we need to seek for possible reasons that could smooth the process in that particular case while simultaneously taking into consideration that structural and temporal resolution (and structural interpretation) of a given experimental setup might mask the true complexity of the folding.

In this manuscript we explain insights that have been obtained by computational studies of GQ folding. We also highlight weaknesses of the computational methods. Contemporary computations are not capable of visualizing full folding events of GQs, in contrast to the fastest-folding proteins. However, the computations can be focused on properties of selected sub-regions of GQ folding landscapes (types of structures that can participate in the folding) and their analysis can be used to understand the basic principles of GQ folding. We do not review theoretical studies that are not related to GQ folding and the reader can find elsewhere a general overview of GQ computations [50]. Finally, we compare computational results with the available experimental data. At present, individual experimental studies suggest a wide range of time-scales and mechanisms of GQ folding. Time-scales of days revealed by some experiments are longer than the measurement times in others. For example the bleaching time in some FRET set-ups is a few minutes [51]. The experimental studies differ widely regarding the nature of suggested intermediate states. In contrast to structural studies of the folded state, experimental studies of folding typically deduce the suggested structures indirectly from measured signals; besides the already cited studies, the interested reader can find many important details on the experimental methods used for example in refs. [52-58]. Thus, it is sometimes difficult to unambiguously separate the direct experimental data from an intuitive structural interpretation. In addition, in the absence of unambiguous structural data, interpretations of experiments may be biased by models currently favored in the literature. Despite the undisputable limitations and weaknesses of computational methods, they indirectly predict that the folding process may involve a very rich spectrum of sub-states, especially in the early stages of folding. If this prediction is correct, then the experimental resolution of so many structures would be difficult, because different conformational basins (even kinetically unrelated ones) may overlap in terms of the measured signals [59, 60].

Among the experimental studies, the one that is most straightforwardly comparable with the computations is a time-resolved structural NMR study (at 298 K) on folding of the human telomeric GQ DNA [13]. It revealed a kinetic partitioning between two basins identified as the hybrid-1 and hybrid-2 GQ folds. The partitioning became visible after ~1.5 hour of real time, and it subsequently took one day to reach thermodynamic equilibrium. It indirectly demonstrates the long life-times of both competing structures (by life-time we denote the time which a single molecule typically spends in a given fold before an excursion into the unfolded state and subsequent either return or transition to another basin). The experiment does not resolve the structural composition of the ensemble in the first 1.5 hour, which may contain substantial populations of some other less stable CBAs with shorter life-times, rather than being truly unfolded.

While our paper was under review, two additional experimental studies appeared [22, 23], which are fully consistent with the preceding MD data [24-27]. They revealed kinetic partitioning (i.e., multi-pathway, branched) model of GQ folding with dominant role played by well-structured four-stranded intermediates. Marchand and Gabelica [23] used electrospray mass spectrometry and CD experiments to study folding of various human telomeric and c-myc GQ sequences. In the initial stages of folding, they found significant populations of off-pathway ensembles with one bound $K^+$ ion that likely corresponded to antiparallel GQs with two quartets, possibly with basket type topologies. This is consistent with our predictions of off-pathway structures with reduced number of quartets and formal strand slippage [24, 26, 27]. The folding landscape by Marchand and Gabelica assumed presence of two two-quartet and four three-quartet ensembles, primarily interconverting via the unfolded ensemble using four kinetically separated pathways. Aznauryan et al. [22] monitored the time course of folding via single-molecule FRET combined with continuum-solvent MD simulations and analyzed with the help of hidden Markov state modelling. They suggested a folding landscape consisting of one two-quartet and three three-quartet ensembles, with all transitions going through the unfolded state. Qualitatively, despite differing in details, all these studies are mutually consistent and agree with the suggestions deduced from MD data [24-27]. The only difference is that we have tentatively proposed an even larger number of competing four-stranded structures than two [13], four [23] and six [22], for reasons explained in ref. [24] and below. However, while MD is capable to identify types of structures having suitable properties to be the dominant competing ensembles, it does not yet allow a quantitative description of the folding landscape. On the other side, the full spectrum of structures may be unresolvable due to the time and structural resolution limits of the experiments, especially in the early stages of folding.

**1.2 The idealized DNA GQ unfolded ensemble and partitioning of *syn–anti* substates.**

Under given conditions and in thermodynamic equilibrium, individual molecules in an ensemble are Boltzmann distributed over the whole conformational space. A complete description of the system is provided by its partition function. Based on the ergodic hypothesis, monitoring a single molecule for a sufficient length of time gives the same distribution of structures as the ensemble. Intuitively, folding is a transition from the unfolded state to the folded state, in which the molecule reaches its native shape. The unfolded state is difficult to characterize, and it is often a mixture of a large number of sub-states with very low populations. The folding process may start from non-equilibrium distribution of such molecules and progress towards the folded ensemble in the equilibrium. Alternatively, one can monitor structural transitions of the molecules at equilibrium, i.e., the time intervals that the molecules spend in different conformations (included unfolded ones) and transition times between them [1].

The folding process depends on the nature of the unfolded (denatured) states, which in turn depends on the specific physical process that has led to unfolding [1]. Thermally denatured, chemically denatured, and low-entropy force-denatured ensembles are not identical and may give rise to diverse folding pathways. These may lead to natural differences between folding processes of the same molecule monitored in different experiments (or computations). For example, force unfolding leads from one low-entropy state (the NBA) to another low-entropy state (stretched unfolded molecules) while temperature unfolding results in a very complex high-entropy unfolded state. The two unfolded ensembles may lack any overlap [61]. Especially for molecules with such complex conformational space as GQs, we should not expect the existence of a single universal folding process. Rather, the folding should be considered as a set of (potentially quite diverse) processes that arise from an intricate interplay between the intrinsic conformational properties of the molecules, the external conditions, and the starting states.

Computations, in principle, allow controllable and reproducible settings of the starting state and of the external conditions. This may simplify understanding of the intrinsic conformational properties of the molecules. For GQs, we have proposed an idealized unfolded ensemble, which is a hypothetical model ensemble of essentially straight single-strand DNA molecules with an equilibrium distribution of *syn* and *anti* populations of the guanines [24]. This is justified by the clear relationship between *syn-anti* guanine distributions and possible GQ folds [62].

We have modeled the idealized unfolded ensemble of 5'-GGG-loop-GGG-3' hairpin DNA sequences in a recent study, using an ensemble of 64 single-strand molecules adopting B-DNA-like conformation and each having a distinct *syn-anti* G combination [24]. Six guanines can adopt 64 unique *syn-anti* patterns. Twelve guanines, corresponding to full three-quartet GQ (5'-$G_3$-loop1-$G_3$-loop2-$G_3$-loop3-$G_3$-3'), can adopt 4096 unique *syn-anti* patterns (Figure 2). Out of them, 2336 are in theory able to fold into a two- or three-quartet GQ. An important result of our T-REMD simulations was the observation that unstructured GQ-forming DNA sequences would populate a wide range of different *syn-anti* patterns, which then have propensities to fold into different GQ topologies [24]. This natural property of the GQ-forming DNA sequences is likely one of the key contributors to the KP mechanism of DNA GQ folding and is often not taken into consideration when proposing simple folding models.

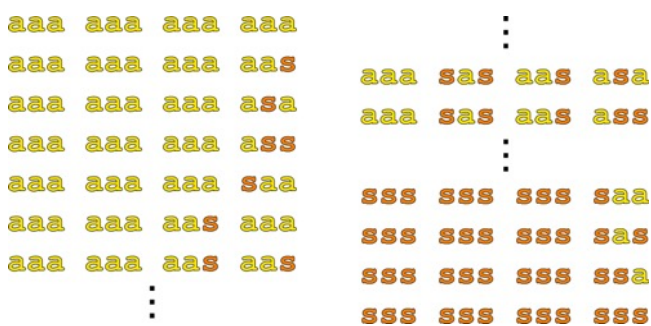

**Figure 2**. Twelve guanosines can adopt $2^{12}$ = 4096 distinct combinations of *syn-anti* pattern. The lines schematically depict the possible *anti* (yellow *a*) and *syn* (orange *s*) conformations ordered from all-*anti* to all-*syn*.

We emphasize that this idealized unfolded ensemble is not *a priori* superior to other possible unfolded ensembles (see above). However, it simplifies, as a model reference state, understanding of the intrinsic conformational properties of the DNA sequences. This ensemble could be related to the high-entropy ensemble corresponding to the temperature denaturation. The *syn-anti* patterns of the G-stretches correspond to the genuine partition of the GQ conformations.

## 2. Molecular dynamics methods to study GQ folding
### 2.1 Unbiased (standard) atomistic simulations.

Unbiased explicit-solvent simulation is the simplest possible application of MD [50, 63, 64]. The simulation starts from an exact conformation (i.e., single configuration of all atoms) of the solute molecule and subsequently mimics thermal fluctuations of real molecules. The simulation time (presently ranging from μs to ms) corresponds to real time. The main weaknesses of the method are the quality of the force field and the short time-scale of the simulations. Overviews of the method, written for non-specialists, can be found elsewhere [50, 63]. MD can be likened to a hypothetical single-molecule experiment initiated from a single conformation (xyz geometry) of a molecule, with essentially unlimited temporal and structural resolution. The best way to correctly interpret the simulation outcome is to imagine what a single real DNA molecule would do in a hypothetical experiment with conditions equivalent to the MD setup, with the same starting structure, and observed with an infinite-resolution microscope for a time corresponding to the length of the MD simulation. The outcome of simulations critically depends on the quality of the starting structure, since the currently affordable simulation time-scales typically do not allow one to repair larger irregularities in the starting structures [65, 66]. With a flawless force field, the simulation would exactly mimic thermal motions of real molecules from the initial configuration. With a flawless force field and unlimited simulation time, the method would provide a converged description of the dynamics of the studied molecule, ultimately leading (according to the ergodic hypothesis) to an equilibrium ensemble population that parallels the one observed experimentally.

Unbiased simulations are the "gold standard" in the field. Nevertheless, their utility in GQ folding studies is limited by the short time-scales that can be simulated with current hardware and software. When starting the simulation in the folded state and using an appropriate force field, the cation-stabilized GQ stem is structurally exceptionally stable [50, 67] and only the loops show local dynamics [25, 68, 69]. The barriers separating any folded GQ from other parts of the conformational space are so large that to see any unfolding event in simulations is extremely improbable. When starting standard simulations from some unfolded state, they are too short to show a significant folding of the molecules. Nevertheless, standard simulations can be used as a supplementary tool to refine description of potential intermediates of the GQ folding and unfolding proposed using other methods (such as enhanced-sampling methods or coarse-grained simulations; see section 2.5) [13, 24-27, 70, 71].

Note that with inappropriate force field parameters, the standard simulation may be unable to keep a full set of ions (one monovalent ion per each inter-quartet cavity) inside the GQ stem. Then a richer dynamics of folded GQs can be observed. However, such ion departure from a G-stem is a sign of severe force field imbalance. Such simulations can be occasionally found in the literature [72], though we suggest that sampling enhancement caused by an inaccurate force field model does not guarantee physically valid results.

2.2 **Relative stability of different GQ stems.**

In a study relevant to GQ folding, standard simulations complemented by continuum solvent calculations estimated relative free energies of different *syn-anti* patterns of GQ stems that contribute to the KP [73]. Specifically, the calculations derived relative free energies of different dinucleotide GpG steps (a*nti-anti, anti-syn, syn-anti* and *syn-syn*) embedded in a GQ stem. Dinucleotides are basic building blocks from which any GQ stem can be constructed (Figure 3). The original computations [73] were then corrected using advanced quantum chemical (QM) calculations, compensating for the limited accuracy of the simulation force field [74]. Due to their computational cost, QM calculations alone cannot be directly used to derive the stability of different conformers/topologies of nucleic acids, since single-structure energy computations do not allow any physically meaningful estimation of free energies (estimation of biomolecular stabilities from single-structure calculations is a widespread misapprehension in the QM literature) [75]. The QM calculations can, however, be used for potential energy corrections of free energies estimated using atomistic simulations, to combine sampling from simulations with the accuracy of QM. The final prediction suggested that the DNA GQ stems in equilibrium tend to maximize the number of essentially isoenergetic *anti-anti* and *syn-anti* GpG steps and minimize the number of *anti-syn* and *syn-syn* steps [74]. Further, if the experimental sequence starts with a 5′-terminal G, strong intra-residue H-bond between the 5′-OH terminal group and G(N3) atom flips the 5′-terminal guanine into *syn* (Figure 3c)*,* unless the 5′-OH group is fully involved with other interactions, for example due to dimerization, multimerization of mutually stacked GQs or some other hydrogen-bonding. These findings are in agreement with the basic rules of GQ topological variability [62, 76].

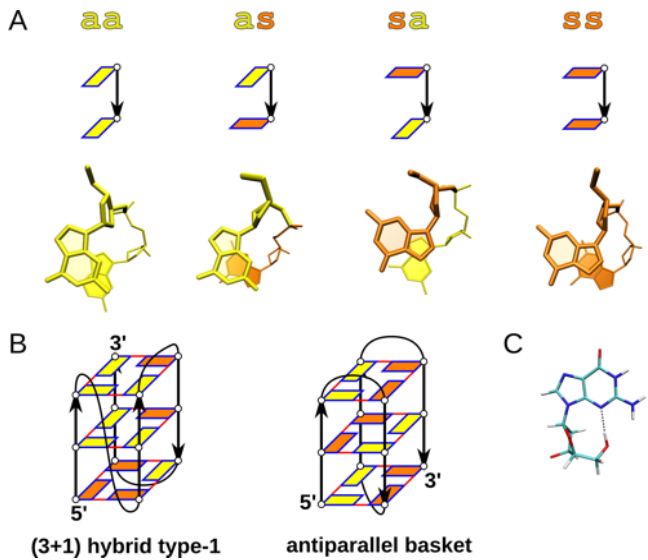

**Figure 3**. Dinucleotide GpG steps. A: *anti-anti*, *anti-syn*, *syn-anti* and *syn-syn* steps (*anti* – yellow, *syn* - orange), the arrow is in the 5' to 3' direction. B: The hybrid type-1 GQ is composed of three *anti-anti* steps, four *syn-anti* steps and one *syn-syn* step; the basket-type GQ contains four *anti-syn* steps and four *syn-anti* steps. C: *syn*-specific 5′-terminal H-bond between N3 and O5′.

**2.3 No-salt simulations.**

Various methods that attempt to speed up structural rearrangements have been applied to GQs and their sub-states [24-27, 70, 71, 77-82]. All these methods have some advantages and also some limitations compared to the unbiased simulations. A technique specific for GQs is to employ no-salt simulations [25, 27]. This approach consists of an unbiased simulation with complete removal of all counter-ions (including those from inside the stem) from the simulation box. The net charge is neutralized by uniformly distributing a compensatory charge over all particles in the simulation box, mainly the thousands of water molecules. No-salt simulations starting from a folded GQ can initiate unfolding of the structures. The process is believed to be relevant to the late stages of folding and early stages of unfolding. Re-folding can then be probed by starting standard simulations with ions from partially unfolded conformations obtained by no-salt simulations, though the refolding attempts are already limited to a small portion of the conformational space accessible using plain MD [27].

The no-salt simulations have a physical justification. The simulation protocol resembles a hypothetical reversal of the stopped flow experimental setup commonly used in studies of GQs. The unfolding initiated by ion-deficiency can be relevant to real unfolding processes, since it is likely that at least part of the real single-molecule unfolding events occurs during periods when the GQ has a temporarily reduced number of ions inside the stem due to genuine ion-exchange processes with the bulk. The no-salt condition is an extreme case of such cation deficiency. An advantage of no-salt simulation is that it does not introduce any biasing force or pre-determined path (reaction coordinate) of unfolding, and that the imposed chemically denaturing conditions, although extreme, have a straightforward interpretation.

2.4 **Ions are essential to stabilize the GQ stems.**

In the complete absence of the ions the otherwise very stable GQ stems are immediately destabilized [27]. The swift initiation of unfolding processes in no-salt simulations confirms that the ion-quartet interactions represent the decisive stabilizing force of GQs, in line with many experiments [83-85]. Without the ions, the fully paired GQ does not appear to be a significant minimum on the free energy landscape and quartet pairing and stacking are not sufficient to stabilize the GQs (though see section 3.6 for our comment on the d[$G_4T_4G_4$] experiment by Plavec et al. [86]). When ions are initially absent in the stem but present in the surrounding bulk, the GQ stem is again immediately destabilized but it is usually capable of spontaneously and quickly capturing ions from the bulk and regains full stability [87]. In contrast, when there are ions inside the stem, the GQ is perfectly stable in the simulations even in the absence of any ions in the surrounding bulk environment (Figure 4) [25]. This indicates that bulk ions are not necessary to stabilize single folded GQs. The GQ fold would lose its integrity only after the ions depart from its channel and the stem is not capable to replace them due to their low concentration in the bulk, which is a very rare event on present simulation times-scales.

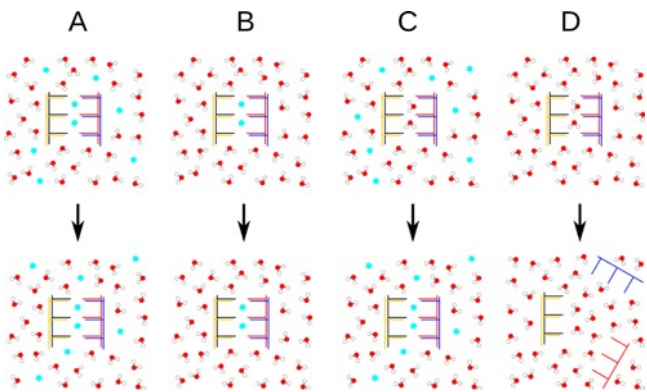

**Figure 4.** Initial ionic conditions in GQ simulations and their consequence. A: Standard simulation with ions in the bulk and inside the GQ channel does not lead to any significant change in the structure of the GQ [50]. B: Cations only present inside the channel are sufficient to keep the GQ stable during the simulation even in the absence of ions outside [25]. C: If no cations are present in the channel, but sufficient concentration of them is in the bulk, the GQ is able to take them up and stabilizes itself [87]. D: In no-salt simulation, the GQ cannot take up any cations from the bulk and ultimately disintegrates [27].

Note that by the word "stability" in the context of MD simulations we usually mean the life-time of a specific conformation of a single molecule adopted at the beginning or during the simulation. When the life-time is much longer than the simulation time-scale, the molecule is called stable. Thus, a conformation that appears as stable in an MD simulation might be metastable on the experimental time-scale.

**2.5 Enhanced sampling simulation methods.**

In order to evaluate thermodynamic stability of different states, MD simulations would have to achieve equilibrium sampling, i.e., converged relative populations of the states. However, based on Eyring equation and first order kinetics, standard simulations can cross the following activation free energy barrier $\Delta G^\ddagger$

$$\Delta G^\ddagger = -RT \ln\left[-\frac{h \ln(1 - p_{\Delta t})}{k_B T \Delta t}\right]$$

where $p_{\Delta t}$ is the probability of crossing the free energy barrier within time-scale $\Delta t$ at temperature $T$. $k_B$, $R$, and $h$ denote Boltzmann, universal gas, and Planck constants, respectively. Thus assuming some reasonable value of transition probability above 50%, the simulations can cross the barriers as low as ~8 kcal/mol and ~12 kcal/mol on 100 ns and 100 µs time-scales, respectively. Thus, natural MD time-scales are often several or many orders of magnitude shorter than typical transition times for interesting events. This even means that in standard simulations starting from experimental structures, we should not see large and irreversible perturbations of the simulated biomolecular folds or complexes, unless the simulation time reaches time-scales of real $k_{unfold}$ and $k_{off}$ constants [65]. Large structural changes in standard simulations may rather indicate force-field problems or errors in the starting structures.

All the above facts have pushed the community to develop a variety of so-called *enhanced sampling* methods. In these methods, the dynamics of the system is artificially altered so as to allow rare (slow) events to be observed in the short time-scale accessible by MD. Ideally, enhanced sampling simulations can then be analyzed to re-construct equilibrium populations (and thus free-energy landscapes) compatible with those that one would have obtained with a much longer standard MD simulation. Different classes and numerous variants of enhanced sampling methods have been proposed and applied with varied success (for a review see e.g. ref. [88]). The major difference among them is the amount of information that is *a priori* required to run the simulation. Properly applied methods can provide striking insights. However, when not properly interpreted, these methods can generate free-energy surfaces and populations that are difficult to be related with experimental data. The same holds for unbiased MD simulations, however, the complexity of enhanced sampling methods makes their assessment more difficult. As we will explain below, the folding landscape of typical GQs is so complex that its *complete* description is certainly out of the reach of any of the existing enhanced sampling methods. We also note that enhanced sampling methods do not reduce in any way force-field deficiencies; in fact, when used properly they unmask force-field imbalances.

### 2.5.1 Enhanced sampling methods based on replica exchange MD.

A widespread enhanced sampling method is "parallel tempering" (PT) [89]; the equivalent and more frequently-used term is "temperature replica exchange MD" (T-REMD). This method minimally requires *a priori* knowledge but is very expensive. In T-REMD, several (typically a few dozens) of replicas of the system are simultaneously simulated at different temperatures (Figure 5). Temperatures are chosen in a ladder where the lowest step represents the physical condition (i.e. room temperature) and the highest step corresponds to a temperature large enough for all the relevant energy barriers to be crossable in an affordable simulation time-scale. The highest temperature is often much larger than any meaningful experimental temperature, even well beyond the boiling temperature of water. This is possible when MD is performed at constant volume. In addition, the molecular mechanics force fields do not allow covalent bond breaking, so the molecules do not decompose at high temperatures. From time to time, typically with a pace of a few ps, a Monte Carlo procedure is used to swap coordinates between adjacent replicas. These swaps are proposed and accepted in such a way that they follow the thermodynamic principle of detailed balance. In other words, an exchange is only accepted when, due to natural energy fluctuations, the conformation from the simulation performed at a given temperature is also compatible with the Boltzmann ensemble at the neighboring temperature. The procedure results in a sort of annealing, where continuous trajectories perform a random walk in the temperature ladder. Continuous trajectories can be followed traveling throughout temperature space, but also discontinuous trajectories can be observed at given temperatures. The later ones converge towards thermodynamic equilibrium of populations of different sub-states at given temperatures. The traveling through the temperature ladder is, unfortunately, partially hiding the true kinetics of the system. When a trajectory reaches the highest temperatures of the ladder, high-energy barriers are crossed. When the same trajectory diffuses to the lowest step of the ladder, the conformation is automatically and progressively adjusted so as to be compatible with the experimental conditions. Usually only the lowest temperature replica is analyzed, as it is the one that is compatible with the experimental conditions. However, it is also important to verify that the continuous trajectories sample compatible ensembles (see e.g. refs [90, 91]). It is in principle possible to also gather the data from the higher temperature replicas and compare them with experiments with different temperatures. However, we caution that while the current force fields are definitely far from perfect for room temperature properties, at higher temperatures their reliability might further decline.

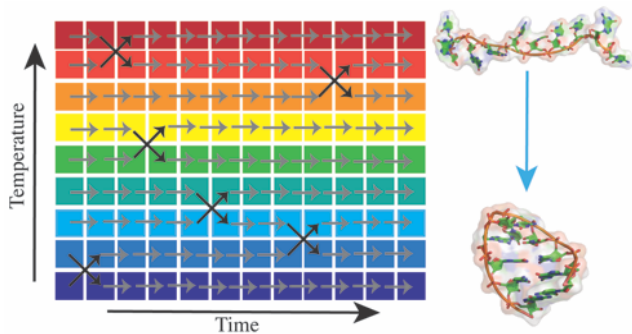

**Figure 5**. T-REMD simulation scheme. Several replicas (separate simulations) are run simultaneously at different temperatures, with periodic exchange attempts. The lowest temperature is typically the experimental temperature and its replica contains folded structures, while the highest temperature should allow crossing of all relevant free-energy barriers. The exchanges are done in such a way that the system should sample Boltzmann distribution at each temperature after convergence is reached.

The basic ingredients of a T-REMD simulation are: a reference (low temperature) replica, which is expected to sample the Boltzmann distribution associated to the force field; a high temperature replica, which is expected to be ergodic so as to sample all relevant sub-states in a statistically converged manner; a number of replicas in the middle with intermediate temperatures. The communication among all replicas guarantees convergence to the correct Boltzmann distribution at each temperature. The number of replicas depends on the system size.

T-REMD is straightforward and requires basically no additional information except a qualified estimate of the size of the typical energy barriers. However, the method is extremely expensive since it often requires tens or hundreds of simulations to be simultaneously performed. Moreover, the efficiency of T-REMD in crossing entropic barriers has been debated [92]. In practice, converged sampling of T-REMD in nucleic acids oligomers has been achieved to date only for RNA tetranucleotides with a huge computational effort [11]. For RNA tetraloops, convergence has been achieved only upon artificially restraining their stems to remain fully paired throughout the entire simulation [11].

T-REMD simulations allow one to access time-scales that are very difficult to observe in unbiased MD. This however could paradoxically make the simulation more difficult to converge. For instance, to converge the local fluctuations around the native structure it is sufficient to run a properly initialized unbiased MD. However, to characterize the folding landscape it is necessary to converge an ensemble that contains both unfolded and folded structures. This can only be done reliably by monitoring that unfolding and folding transitions are observed in the continuous trajectories in a statistically significant manner [12].

Since there is no guarantee that the global free energy minimum associated with the force field coincides with the native structure, a strict validation of both the force field and the convergence of the simulation can be obtained only by initializing T-REMD simulations in an unfolded structure [93]. Conversely, applications where short T-REMD simulations are initialized in the native structure can only provide information on the local flexibility of this specific conformation, similarly to unbiased MD, with the disadvantage of sampling conformations corresponding to an unknown temperature. Such studies [94] should not be interpreted as folding studies though they may provide some idea about unfolding pathways under thermal denaturation.

We have recently applied T-REMD (64 replicas, temperature ladder 278-445K, 1 μs per replica) to the folding of a human telomeric DNA GGGTTAGGG G-hairpin, using an idealized unfolded ensemble (see section 1.2) to initialize the simulation [24]. This has been so far the only T-REMD simulation study relevant to the GQ folding that has been started from the unfolded state. The convergence of the simulation was still not enough to *quantitatively* estimate the relative stability (populations) of the sampled conformers. However, this simulation achieved an exhaustive sampling of the G-hairpin conformational space so that it could be assumed that no significant structures from the conformational space have been missed. It provided useful insights into events relevant to the earliest phases of the GQ folding as well as to later GQ structural transitions involving folding and unfolding of G-hairpins (see section 3.1).

The replica exchange protocol can be generalized to methods where ergodicity is not obtained by increasing temperature, but by either scaling portions of the force field [95, 96] or adding penalty potentials disfavoring specific structures, for example, biasing selected dihedral angles [97, 98]. These methods are known as Hamiltonian replica exchange (H-REMD) methods, since the different replicas follow different Hamilton equations. Again, the unbiased replica is used to gather the data to be compared with experiments.

### 2.5.2  Enhanced sampling methods using collective variables.

A blind acceleration of the whole simulated system by raising its temperature in T-REMD might be a waste of resources in cases where some *a priori* information about the most important conformational bottlenecks (transition states) on the folding landscape is available. Methods based on collective variables (CV) usually allow encoding such *a priori* information in the simulation protocol and may then provide a much more efficient computational tool than T-REMD. CV-based methods represent the second branch of enhanced sampling methods.

CV-based methods are essentially projecting the free energy landscape from the full multi-dimensional coordinate space onto a few selected generalized (reaction) coordinates, allowing the definition of a low-dimensional free-energy surface. The name collective variable reflects that CVs aim to describe a collective motion of the system. The CVs are chosen in order to include the most relevant dynamics (slow motions) of the studied processes by assuming that the rest of the dynamics are either unimportant for the studied process or are just thermal fluctuations that can equilibrate on a short time-scale. The root of CV-based methods is umbrella sampling [99]. Here, a suitable degree of freedom should be chosen to effectively describe the slow modes of the system. Then, a biasing potential is added on this degree of freedom so as to accelerate sampling along it. For instance, in the case of an isomerization process, a biasing potential acting on the relevant torsion could be sufficient. Ideally, such an additional potential should stabilize the transition state so as to increase the probability of observing transitions, in a fashion similar to a catalyst. The biasing potential aims to flatten the probability distribution along the torsion. The potential can then be discounted *a posteriori*, and the natural unbiased probability distribution (unbiased free energy surface) can be straightforwardly reconstructed from the biased one, provided the simulation is converged.

Several methods to build penalty potentials iteratively have been proposed. In these methods, the penalty slowly builds up from zero with the simulation time based on the preceding simulation behavior. The aim is to flatten the probability distribution along the CVs (Figure 6). One of the most popular methods is metadynamics [100], where a history-dependent procedure is used to disfavor already visited conformations. The penalty potential is constructed as a sum of Gaussians. In its well-tempered variant (WT-metadynamics), the speed at which the penalty potential grows decreases during the simulation, reaching a quasi-equilibrium state [101]. Moreover, in WT-metadynamics one can easily tune the parameters of the simulation to obtain a partial flattening of the free-energy landscape, avoiding the exploration of unnecessary high-energy states. The width of the Gaussian can also be chosen automatically [102]. WT-metadynamics has been applied to the characterization of the GQ folding landscape in ref. [77].

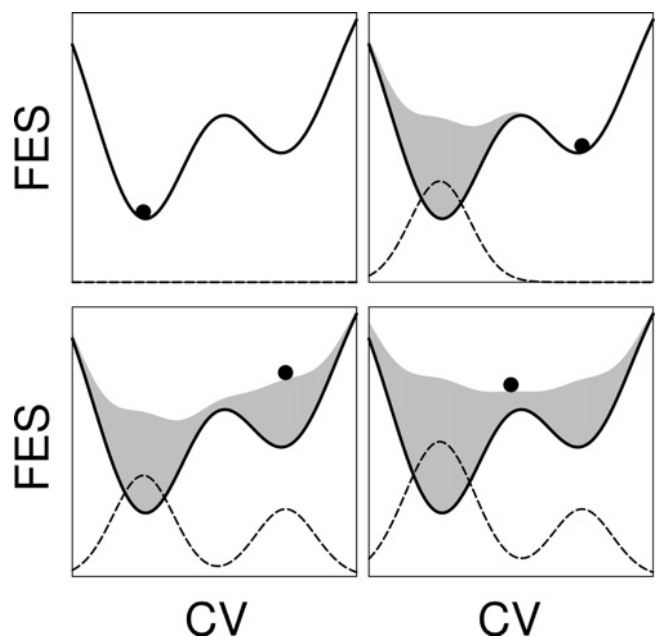

**Figure 6**. Cartoon illustrating how metadynamics samples a double-well free-energy landscape (black solid line). A growing penalty potential (dashed line) is constructed iteratively by adding Gaussians onto already visited values of the CVs. The free-energy landscape felt by the simulation (a sum of the parent potential – the solid black line – and the added bias potential – the grey area) is the upper boundary of the gray area. The simulation is initiated in the left basin (top left). As soon as this basin has been filled by the penalty potential, the system displays a transition to the second basin (top right) and the simulation starts to build up the penalty potential there (bottom left). When both basins are filled, the system can freely diffuse along the CV coordinate (bottom right). The biasing potential will then be approximately equal to the negative of the parent potential. Note that the method works in the space of the collective variable and assumes that dynamics which is not included in the collective variable (it is orthogonal to the CV) is unimportant. See the main text for explanation of the difference between metadynamics and WT-metadynamics.

In all CV-based methods a crucial role is played by the choice of the accelerated degrees of freedom, i.e., CVs. Note that even when using the "history-dependent" methods such as metadynamics, the outcome is ultimately determined by the chosen CVs. A simulation performed using ineffective CVs is very difficult to converge and may entirely miss-interpret the actual free energy landscape. Examples of how to detect erroneous CVs in metadynamics can be found in refs. [103, 104]. On the other hand, when appropriate CVs are used, methods of this class are very effective and allow for a speed up of several orders of magnitude compared to standard MD [105]. Analytical methods to recover not only the unbiased population distribution but even the unbiased rate constants have been recently proposed [106]. The search for CVs suitable for describing a given process is often based on a trial and error procedure. For this reason, the community of users of these methods has developed a large number of CVs suitable for different systems. The software package PLUMED implements a large number of CVs, and allows CV-based enhanced sampling methods to be used with a variety of MD codes and in a wide range of applications [107]. Simple CVs are for example RMSds from certain target structures or the radius of gyration of a molecule. Equation (1) illustrates a more complex CV ($CV_{H\text{-bonds}}$) that can be used to monitor the number of native H-bonds in the course of the simulations. The summation is done over all of the H-bonds and $r_i$ are the individual H-bond distances. For systems with N native H-bonds, the value of $CV_{H\text{-bonds}}$ approaches N in the folded state and 0 in the fully unfolded state. The parameters $r_o$, and $n$ are used for a smooth switching between presence and absence of the H-bonds. Typical $r_0$ and $n$ values are ~2.0-2.5 Å (for the H…X distance) and 6, respectively. Smoothing functions are required to allow the resulting CV to be a continuous (differentiable) function of the atomic coordinates. It is important because the penalty potential should be translated into forces acting on the atoms in MD.

$$CV_{H-bonds} = \sum_i \frac{1}{1+\left(\frac{r_i}{r_0}\right)^n} \qquad (1)$$

The CVs can be intricate functions of the atomic (x,y,z) coordinates of the studied molecules. In the case of $CV_{H\text{-bonds}}$ all $r_i$ values are calculated from the atomic coordinates. Note that $CV_{H\text{-bonds}}$ merges all the individual H-bonds (for example, all the native H-bonds in the GQ stem) into one super-coordinate. This leads to the desirable dimensionality reduction of the description but also may lead to oversimplification. The $CV_{H\text{-bonds}}$ reflects how many native H-bonds are present in a given simulation snapshot, but it does not differentiate which of them are present. If there are different long-lived kinetically separated sub-states with similar fractions of formed H-bonds they would overlap in the $CV_{H\text{-bonds}}$ projection. The principal limitation of $CV_{H\text{-bonds}}$ for GQs is illustrated by Table 1, which shows that the different known folds of human telomeric GQ have very diverse sets of native H-bonds. Thus, e.g., when using the 24 H-bonds of the basket GQ topology as the native state to define $CV_{H\text{-bonds}}$, all other known folds would have only 2 - 8 "native" H-bonds formed. $CV_{H\text{-bonds}}$ would count them rather as part of the unfolded state, since even a properly folded G-hairpin may contain 6 native H-bonds. Thus, $CV_{H\text{-bonds}}$ can be efficient to describe simple folding of molecules where the number of native H-bonds progressively increases along the folding pathway. However, it is inadequate to describe transitions between the different major basins on the GQ folding landscape. This example shows that whereas the number of native contacts can be fruitfully used to monitor the progression along the folding pathway in a funneled landscape [5], it is not a good variable in a system with a KP landscape. Note that Tsvetkov et al. recently suggested some alternative set of CVs designed to describe conformational behavior of GQs [108]. These CVs capture various deformations of GQ, which might be useful in analyses of structural dynamics of more-or-less stable structures. However, they do not allow the description of topological rearrangements, since they suffer from the same problem as the $CV_{H\text{-bonds}}$ coordinate.

**Table 1**. Number of GG H-bonds that are shared by distinct folds of the human telomeric GQ (flanking residues are not considered). The Table illustrates that it would be very difficult to propose a CV based on the number of native H-bonds that would be suitable to study transitions between different GQ folds.

|  | Parallel (PDB ID: 1KF1) [109] | Hybrid type-1 (e.g. PDB ID: 2GKU) [38] | Hybrid type-2 (PDB ID: 2JPZ) [33] | Basket (PDB ID: 143D) [110] | Antipar. w. prop. loop (PDB ID: 2MBJ) [111] | Two-quartet (PDB ID: 2KF8) [112] |
|---|---|---|---|---|---|---|
| Parallel | 24 | 12 | 12 | 2 | 12 | 2 |
| Hybrid type-1 | 12 | 24 | 12 | 4 | 8 | 4 |
| Hybrid type-2 | 12 | 12 | 24 | 4 | 8 | 2 |
| Basket | 2 | 4 | 4 | 24 | 6 | 8 |
| Antip. pr. l. | 12 | 8 | 8 | 6 | 24 | 2 |
| Two-quartet | 2 | 4 | 2 | 8 | 2 | 19 |

Replica exchange and CV-based methods can be combined. A typical example is parallel tempering metadynamics [113], where multiple metadynamics simulations at different temperatures are performed and combined. Another approach is bias-exchange (BE) metadynamics [114]. In this approach (not using temperature acceleration), several replicas are simulated, each with independent metadynamics simulation using a different CV. The difference from conventional metadynamics is the following: conventional metadynamics combines several CVs to construct a true multi-dimensional description of the system in the space of the used CVs. However, the computational requirements increase sharply with the number of the CVs, so typically only two or three CVs are used. In BE metadynamics, the individual simulations with different CVs are running independently. This allows one to use more CVs, however, the coupling (inter-dependence) between CVs is not explicitly taken into consideration and the penalty potentials act on one CV at a time. If such penalty potentials are sufficient to induce an exhaustive exploration of the conformational space, the free energy landscape can then be reconstructed, usually as a function of a few of the biased CVs. BE metadynamics has been used to visualize potential unfolding pathway of a hybrid topology of the human telomeric GQ [70].

The free energies in MD papers are usually presented as plots of free energies or populations as a function of the CVs. To get free energy values of states comparable to experiments, it would be needed to split the whole plots into the substates and integrate the populations over their respective subspaces. When such integrated values are not reported, and if the substates are associated with clear minima in the free-energy plots, meaningful estimates of the free energy differences can be obtained by just considering the differences between the minima. A general warning about evaluation of the free-energy plots is that all the relevant states contributing to a basin should be sufficiently sampled. When this is not the case, the result could be affected by sizeable statistical errors. Note that other CVs can also be a posteriori analyzed through a reweighting procedure, provided that the biased simulation captures the essence of the studied process [77, 115, 116]. In other words, one can obtain the free energy as a function of CVs different from those originally biased. In converged free-energy calculations, this post-processing analysis allows one to calculate multidimensional free-energy profiles, which are very useful in studying complex phenomena such as protein/DNA folding.

### 2.5.3 General limitations of enhanced sampling methods for GQs.

As we have pointed out elsewhere [24, 49], it is not yet clear whether the CV-based methods can be used to describe the full GQ folding landscape. CV-based methods can be successfully applied only when the chosen set of CVs is sufficiently complete to describe all relevant motions of the systems (i.e., slow modes of dynamics). For more complex systems, it may be difficult to find the right CVs. For some systems such "coarse-graining" of the free energy landscape may even be virtually impossible due to its natural complexity. For example, it is possible to find suitable CVs to initiate unfolding of a GQ from a given topology. However, it is unclear how to simultaneously describe, using a small set of CVs, all the different GQ folds and intermediates present in the full GQ folding landscape, i.e., to achieve transitions between the different folds. Different free energy basins may dictate differently constructed CVs, as illustrated above for the $CV_{H-bonds}$. It may be equally difficult to couple the folded basins with the unfolded state ensemble, since for GQ folding any CV-based description would have to, among other things, accelerate uncorrelated sampling of all *syn* − *anti* transitions of all guanines [24], which appears to be a demanding challenge in the CV description.

In summary, enhanced sampling methods can provide striking insights, but are not a panacea. These methods, when properly applied within their genuine applicability limits, can dramatically speed up and visualize key structural developments in the simulations. On the other hand, when not applied wisely, these methods can lead to over-interpretations. Due to the complexity of the methods, it may be difficult for readers to assess the relevance of studies lacking an explicit discussion of the limits. Note that construction of CVs is the most fundamental issue for all CV-based methods (metadynamics, steered MD [117], umbrella sampling, adaptive biasing force [118],…) irrespective of their specific implementations. Thus, readers can first try to assess sufficiency (completeness) of the chosen set of CVs, before analyzing other details of the method. Common sense can often indicate if the method is already beyond its genuine applicability limits. The easiest assessment can be made by checking if a CV is truly capable of describing and distinguishing the relevant stable or metastable conformations. Another consistency check can be made by looking at the reported free-energy landscapes. In CV-based method, the maximum boost is of the order of $\exp(\Delta G^{\ddagger})$. The folding of human telomeric DNA GQs often takes days, which is ~11 orders of magnitude longer than the time-scale affordable by MD. Only a bias potential that favors the probability of the transition state by approximately 15 kcal/mol could lead to such a boost.

Moreover, similarly to unbiased MD simulations, all enhanced sampling simulations (both T-REMD and CV-based methods) should report multiple events to be statistically significant. This for human telomeric DNA GQs brings the following requirements: first, repeated crossings between a given fold and the unfolded states; second, full sampling of the unfolded state with all its 4096 guanine *syn-anti* sub-states; third, converged crossings between the different competing GQ folds. Thus, some studies that might be misinterpreted as a full characterization of the GQ folding in fact could only characterize some sub-pathways of the whole process. On the other hand, when properly commented and interpreted, even a partial description of folding pathways might uncover unique information that is not accessible with standard MD. To the best of our knowledge, full folding of GQs is beyond any of the currently available computational tools.[49]

### 3. GQ folding intermediates
### 3.1 G-hairpin intermediates as diverse fast-folding species.

It is widely accepted that G-hairpins form very rapidly and represent the first stage of folding of DNA GQs [15, 119-121]. Hairpin-like structures may participate also in many transitions in all subsequent stages of GQ folding. MD simulations suggest the following properties of G-hairpins [24].

i. Antiparallel hairpins fold easily on the microsecond time-scale, but no clear preference for formation of any specific *syn-anti* pattern (with respect to native GQ topologies) is observed.
ii. Misfolded species with non-native H-bonding and with shifted strands are abundant (Figure 7).
iii. Stability of antiparallel hairpins depends on interplay between their *syn-anti* pattern, groove width and the presence of flanking bases (Figure 7), making some of them potent nuclei for further folding into triplexes with lateral and diagonal loops.

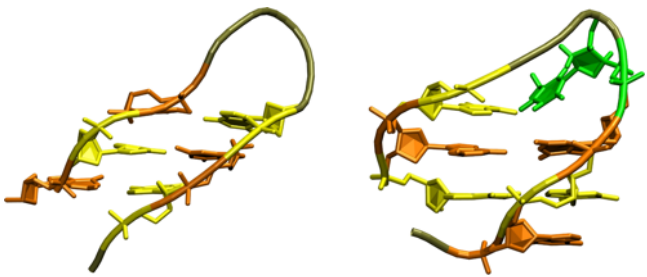

**Figure 7**. Antiparallel G-hairpin d[AGGGTTAGGG] [24]. A stable G-hairpin potent for the formation of the basket-type GQ (left) and a misfolded G-hairpin with shifted strands (right). *Syn* Gs are in orange, *anti* Gs are in yellow. The green thymine in the misfolded structure forms a GT base pair. Other residues or ions are not shown.

The existence of a spectrum of various antiparallel hairpin structures supports the idea of a multi-pathway nature of the folding process from its outset. On the other hand, parallel hairpins with propeller loops are very unstable and transform quickly into the cross-like arrangement [79] or even unfold completely. It suggests that parallel hairpins either do not exist as separate entities and enter the GQ folding process in much later stages, or that the force field is unable to describe them properly.

**3.2 G-triplex intermediates as popular but somewhat enigmatic species.**

G-triplexes are the most popular intermediates in contemporary GQ folding literature. Nevertheless, the exact role of G-triplexes is yet to be elucidated, since the currently available experimental and computational data contain some potential contradictions and uncertainties. Let us explain this issue point by point, in order to reconcile all the available results.

The idea of three-triad G-triplexes participating in folding of human telomeric GQ came from MD simulations and experiments by Sugiyama's group (Figure 8) [122, 123]. However, the simulations were very short; the 3 ns time-scale is not sufficient to judge structural stability of simulated biomolecules as almost everything survives 3 ns simulations. Further, they were executed using the outdated ff99 (see section 4) force field. Longer simulations with ff99 would result in rather unstable trajectories. Nevertheless, rather reasonable stability (life-time) of the G-triplexes was then confirmed by μs-scale simulations with refined force fields [26]. In addition, formation of G-triplexes has obtained indirect support from a series of elegant non-atomistic experiments [124-126].

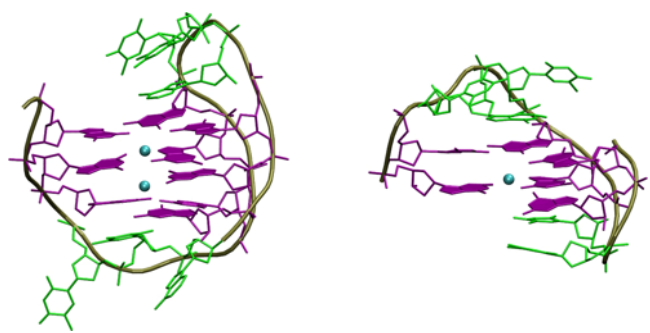

**Figure 8**. Three-triad G-triplex d[GGGTTAGGGTTAGGG] (left) and two-triad G-triplex d[GGTTGGTGTGG] (right). G-stem is shown in mauve, loop residues are in green, channel cations are in cyan.

Formation of a two-triad G-triplex has also been predicted for folding of the 15-TBA (thrombin binding aptamer) DNA GQ using WT-metadynamics and the bsc0 AMBER force field (Figure 8) [77]. This theoretical prediction has been accompanied by unambiguous experimental evidence (NMR, DSC and CD experiments) of formation of such a triplex using an appropriate truncated 15-TBA construct [77]. In this work, a WT-metadynamics simulation biasing the gyration radius and the number of hydrogen bonds in the G-stem was reported for the full sequence of the 15-TBA, starting from the folded structure. Though the CVs used are not able to fully describe GQ folding and unfolding, the simulation suggested a G-triplex as a metastable conformation. The WT-metadynamics was supplemented by a 100 ns ordinary MD simulation of the truncated G-triplex, suggesting that it was structurally stable. However, subsequent multiple and longer unbiased MD simulations using the same force field on the same truncated triplex structure revealed that the G-triplex was unstable and typically fell apart on the ~100 ns time-scale [26]. It cannot be ruled out that, besides the different length of the simulations, the opposite results could have been affected by some differences in the simulation protocols.

The later simulations nevertheless indirectly indicated that the two-triad G-triplex was not thermodynamically stable within the description of the bsc0 force field (see section 5.) compared to the unfolded state, as no signs of re-folding were observed [26]. The time-scale of the loss of the folded structure can be used to estimate the $k_{unfold}$ rate constant, while the lack of any sign of refolding in the subsequent much longer parts of the simulations gives a lower limit of the $k_{fold}$ rate constant. More precisely, since the average unfolding time was ~100 ns in bsc0, the corresponding $k_{unfold}$ would be $10^7$ s$^{-1}$. We suggest that it is not compatible with a thermodynamically stable folded molecule, even in absence of a rigorous estimate of the $k_{fold}$. (The time spent in the unfolded state before the simulations were terminated would already indicate $k_{fold} < 2 \times 10^6$ s$^{-1}$, in absence of any refolding events). For example, if this simulation set is used in a typical Markov State Model analysis, the initial state would be removed from the active set of microstates and discarded from the analysis [127]. One of the Reviewers suggested that the absence of refolding could be a kinetic problem. In other words, due to the roughness and high-dimensionality of the free energy landscape in DNA folding, the molecule could be trapped in some specific conformation in the unfolded ensemble, kinetically preventing refolding. Whereas this is possible, we recall that in the discussed set of simulations [26] the unfolded conformations were directly evolved from the folded ones. This is the reason why from these simulations we can only propose a lower boundary rather than an actual estimate for the life-time of the unfolded state. Nevertheless, there was a high flux of the molecules from the starting structure to the unfolded state but there was no sign (on the simulation time scale) of a backward flux, with the simulations progressively evolving away from the starting state. Thus, assuming the validity of the detailed balance (microscopic reversibility) principle, we suggested that the folded state was not thermodynamically stable [26]. Perhaps we cannot exclude the possibility that the starting structure contained some high-energy structural feature not resolvable by standard equilibration protocols. Such internal stress could potentially destabilize the simulated molecule during the early stages of simulations.[65]. Then, the specific starting structure could have a much higher probability to unfold than to fully relax, which could bias the results.

Nevertheless, we reemphasize the unambiguous support for formation of the G-triplex came from experiment, particularly from NMR of the truncated sequences containing three 15-TBA G-stretches [77], irrespective of the diverse simulation results. The results of the longer simulations thus might indicate that the current force fields (bsc0) underestimate the stability of the G-triplex structure (see section 5.2) and we strongly suggest that this is the explanation of the above-discussed results. We, however, fully respect that there could be alternative interpretations of the results. Importantly, a subsequent NMR study of the same authors of ref. [77] has resolved the solution structure of the 3′-end-TBA-truncated 11-mer oligonucleotide (11-mer-3′-t-TBA; PDB ID: 2MKM) at low temperature (274 K). It showed that 11-mer-3′-t-TBA forms two G-triad planes very similar to those predicted by the previous metadynamics simulations, while for other TBA-truncated sequences, G-triplex could not be detected [128]. The different stability of the TBA-truncated sequences and their heterogeneity demonstrated by the CD melting curves supported the role of the length of TT loops on the stabilization of the G-stem.

We have also performed a very extensive set of unbiased MD simulations starting from the folded G-triplexes of the human telomeric sequence [26]. The simulations suggested that the three-triad G-triplexes have life-times typically below 1 μs. This might be too short for major roles in the folding of GQs, unless the G-triplexes have exceptionally fast $k_{on}$, which actually was not indicated by the simulations. In other words, the MD technique suggested that G-triplexes can survive for some time if formed but were not stable in a thermodynamic sense even in the absence of any other competing structures (which would be present in case of full GQ-folding sequences). In line with this, the time-resolved NMR could not detect any G-triplex intermediates during folding of hybrid human telomeric GQs [13].

How to reconcile all the above studies? First, the experimental support by DNA origami experiments and by NMR for the formation of G-triplexes under appropriate conditions appears undisputable [77, 124-126, 128]. Lower stability of the two-triad G-triplex is not necessarily a problem, since three-triad G-triplexes may be considerably more stable [26]. A greater problem is the sub-µs life-time of the three-triad G-triplexes predicted by MD [26]. However, as explained in the subsequent section of this review, we suspect that the force field may considerably underestimate stability of the G-triplexes. Thus, the actual life-times of three-triad triplexes can be longer by a few orders of magnitude compared to the present MD picture [26]. This would make the G-triplex a very competent potential intermediate. Regarding the remarkable time-resolved NMR study by Bessi et al. [13], this experiment does not provide structural insight into the first 1.5 hours of the folding process, which creates a space for participation of G-triplexes in this time window. For the sake of completeness, Aznauryan et al. [22] tried to detect folded G-triplexes using an appropriate construct but no FRET signal that would be consistent with G-triplexes was found.

Summarizing all the data, it is evident that G-triplexes could somehow participate in the GQ folding processes. What remains to be clarified is their kinetic accessibility and life-times with respect to the other types of intermediates. More specifically, if and how they outcompete during specific phases of the folding process various four-stranded intermediates [26]. Ion-stabilized G-quartet intermediates should have considerably longer life-times than G-triplexes. Even when not being major detectable free-energy sub-states on the folding landscape, G-triplexes may be certainly involved in many diffusive transitions between various basins on the folding landscape, keeping a relatively stable core of the molecule for eventual strand rearrangements.

Disregarding the issue of potential underestimation of the G-triplex stability, the MD simulations provided several insights into the structural dynamics of G-triplexes [24, 26]. Those with lateral and diagonal loops were found to have reasonable life-times. The simulations have also revealed that the G-triplexes can easily rearrange between a triplex with two lateral loops and a diagonal+lateral loop triplex (Figure 9). The transition requires one G-stretch to change its binding partner and proceeds via an intermediate triangle-like triplex. Such intermediates contain a spectrum of GG base pairs ranging from Hoogsteen (GQ-like) geometry to a reversed Watson-Crick pair. So, transitions between diagonal and lateral loops are facile to achieve. However, as in the case of parallel G-hairpins, triplexes with the propeller loop are predicted to be unstable and quickly unfold, often via cross-like structures [26].

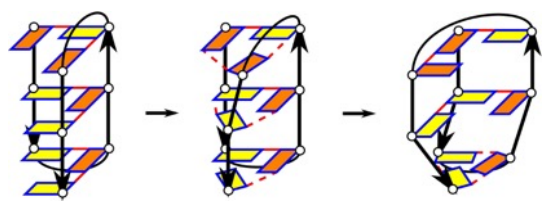

**Figure 9**. Transition of a triplex with two lateral loops into triplexes with diagonal+lateral loop. *Syn* Gs are in orange, *anti* Gs are in yellow.

### 3.3 GQs with shifted strand as potential long-lived intermediates have been overlooked to date.

Alternative (so far unobserved) cation-stabilized four-stranded structures with shifted strands and reduced number of G-quartets may be important off-pathway intermediates [27]. They could cause substantial roughness of the folding landscape, since they have non-native G-strand *syn–anti* conformations and thus cannot be easily rearranged by strand slippage without some degree of unfolding.

Although the role of strand-shifted GQs has so far been largely overlooked in experimental studies, there are several reasons to consider them. First, one of the experimentally detected human telomeric DNA native folds (PDB ID: 2KF8) [112] is such a structure. Thus, it is certainly possible that structures of this type are populated as competing basins of attraction during the folding process of other GQs. Ion-stabilized two-quartet stems should have much longer life-times than the more popular G-triplex intermediates though G-triplexes could be more easily kinetically accessible. From a given three-quartet fold one can form eight alternative two-quartet GQs by a single strand-shift. Shift of two strands can lead to additional six two-quartet GQs (Figure 10). Thus, four-stranded intermediates with shifted strands can be entropically favored over the native structures. As noted above, presence of long-living two-quartet off-pathway intermediates has been supported by the latest experiments [22, 23].

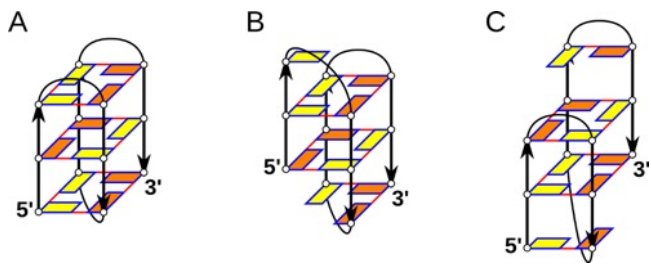

**Figure 10**. GQs with shifted strands. A: Native basket-type GQ [110]. B: GQ with one shifted strand. C: GQ with two shifted strands. The structures in B and C cannot readily transform into the native basket, because they would need to partly unfold and reorient *syn* (orange) and *anti* (yellow) conformation of some guanosines. Note that the structure depicted in B would only need to flip two guanosines in the bottom triad to an *anti* orientation and the native two-quartet GQ (PDB ID: 2KF8) [112] would be then formed.

**3.4 Idealized master pathways vs. a KP mechanism.**

Figure 11 shows a possible folding pathway for the basket topology of human telomeric GQ [24]. It resembles similar pathways derived from experimental studies and depicts the folding progressing rather straightforwardly through a limited set of intermediates. This particular folding pathway has been suggested based on extensive MD simulations of potential intermediates that can participate in the folding, their structural stabilities and transitions between them. It illustrates how MD data can be used to propose simple pathways of GQ folding. However, we have cautioned that such simple intuitive pathways are unlikely to adequately represent real folding processes [24]. The first entirely unrealistic prerequisite is that the *syn-anti* patterns of the approaching components are always the native ones. This common simplification *a priori* ignores interference from off-pathway intermediates with different *syn-anti* patterns, including alternative three- and two-quartet GQs. More likely, such idealized pathways show only one possibility and in the best case could represent 'master pathways', along which other misfolded intermediates occur. Taking into account the slowness of the GQ folding, we suggest that only a tiny fraction of the individual molecules reaches the native basin (fold) directly via the master pathway and the true folding follows the KP mechanism. The idealized folding pathways also do not take into account a non-specific collapse, in which denatured oligonucleotides form 'coils' [1] not resembling the GQ, and then slowly by diffusion and random rare transitions progress towards the native or misfolded GQs.

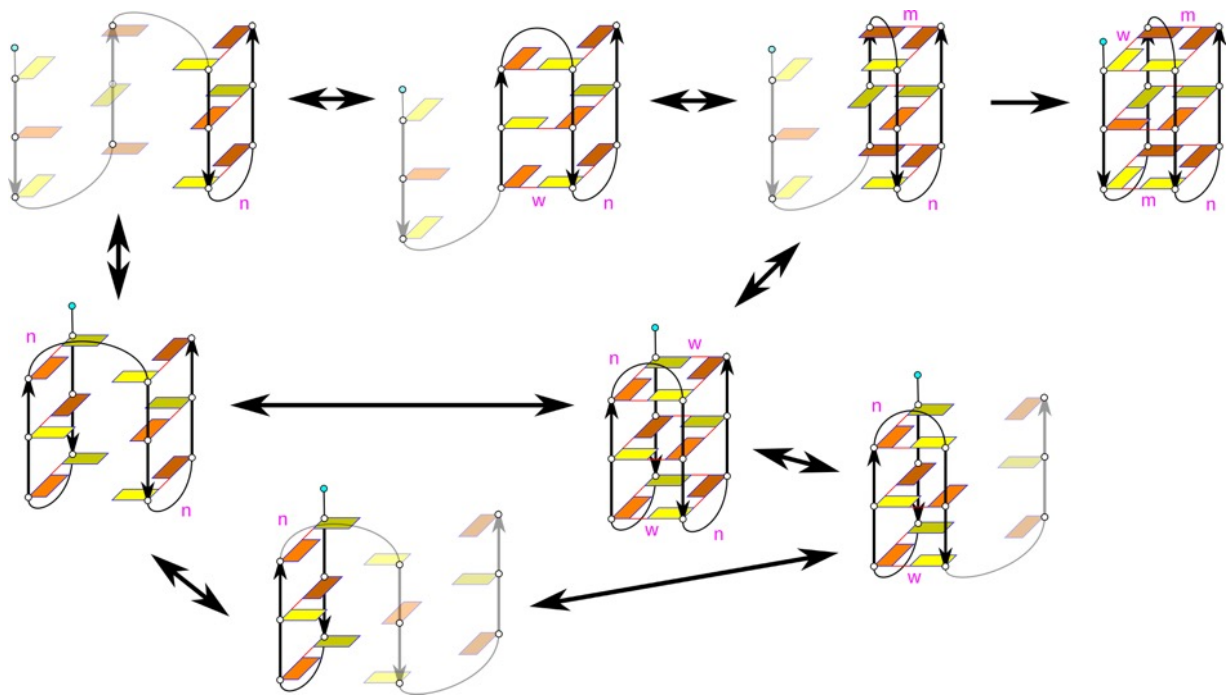

**Figure 11**. Idealized folding pathway of the basket topology. First, G-hairpins are formed. Then the folding proceeds through G-triplexes until it reaches the native basket GQ state (top right). One of the folding branches contains a chair-like GQ. Note that *syn* (orange) ↔ *anti* (yellow) transitions of guanosines are ignored in such pathways.

### 3.5 Formation of parallel-stranded tetramolecular GQs.

Earlier MD simulations were used to propose a potential mechanism of formation of a tetramolecular parallel-stranded GQs [79]. The model has been derived from short simulations with the ff99 force field available at that time and received some experimental support [129, 130]. The simulations suggested that early stages of the formation may involve G-duplexes with imperfectly paired strands (cross-like conformation) while the late stages of the process may involve GQs with strand slippage and reduced number of quartets. When assuming all-*anti* orientation of the guanines, the slipped stems can easily progress towards the native stem with the maximized number of quartets (Figure 12). Once a first G-quartet forms, it is immediately stabilized by monovalent ions. The other intermediates may include three- and four-stranded rather disordered coil-like complexes weakly stabilized by cations. It is likely that modern computations with improved force fields and more than three orders of magnitude longer time-scales could in future further refine this decade-old computational model of formation of tetramolecular GQs.

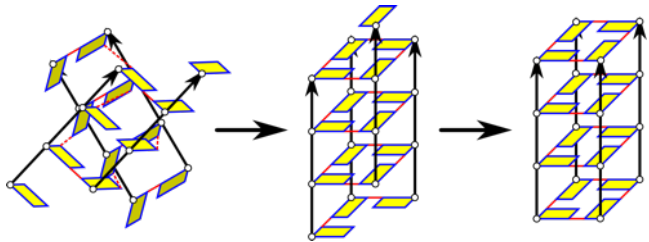

**Figure 12**. Possible late-stage folding of a parallel tetramolecular all-*anti* GQ. Double cross-like intermediate (right) rearranges to a GQ with shifted strands (middle) and one of its strands slips by one level downwards to form a proper GQ (right). Cations are not shown; all the structures should be binding at least one cation to be stable enough for the folding [79].

### 3.6 Number of G-quartets may dramatically affect the folding process.

Most of the above-noted results were obtained for human telomeric GQ DNA sequences. However, the folding process may dramatically depend on the number of Gs in the G-strands, i.e., the number of quartets in the native structures. The 15-TBA GQ can be a specific case, as its folding may be greatly simplified by its inability to form strand-slippage four-stranded structures and thus its limited repertoire of stable competing folds. $G_4$ stretches may also specifically modify the folding landscape compared to the human telomeric GQ DNA as they could stabilize pre-folded structures via stacking and pairing. A recent NMR study for $d[G_4T_4G_4]$ in the absence of ions suggested the formation of a four-stranded $d[G_4T_4G_4]_2$ structure with two weakly interacting diagonal hairpins, which seemed to rapidly convert into the GQ after adding ions (Figure 13) [86]. Such intermediates have to date not been considered and their investigation is desirable as they further increase the complexity of the folding landscape.

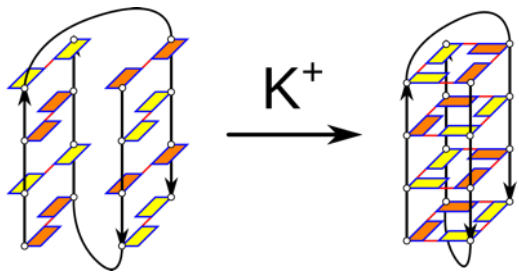

**Figure 13**. Four-stranded intermediate in folding of the *Oxytricha nova* $d[G_4T_4G_4]_2$ GQ in absence of cations proposed by Ceru et. al. [86]. The intermediate (left) binds no structural cations and its GG base-pairs are of the reverse Watson-Crick type. Upon $K^+$ addition, it quickly transforms to the GQ (right). *Syn* Gs are in orange, *anti* Gs in yellow. The $K^+$ ions inside the channel of the formed GQ are not shown.

### 3.7 RNA GQs.

Folding of RNA GQs may differ from DNA GQs because, on the basis of currently known structures, RNA GQs adopt only *anti-anti* GpG dinucleotide steps, which may simplify the kinetic partitioning. On the other hand, interactions of the RNA 2'-OH groups may significantly increase the local ruggedness of the conformational space. There have been so far no computational studies on RNA GQ folding and in addition there is little experimental data.

### 4. Atomistic force fields utilized in studies of GQ folding.

Quality of simulation force fields has a major impact on simulation studies of GQ folding. Essentially all atomistic simulations reported in this work were obtained by different consecutive refinements of the seminal 1995 AMBER Cornell at el. pair-additive NA force field [131]. It is the simplest type of fixed-parameters atomistic force field, modelling covalent structure by harmonic springs for bond lengths and bond angles, supplemented by non-bonded terms consisting of Lennard-Jones van der Waals spheres and atom-centered fixed point charges. The whole parametrization is completed by dihedral potentials, which are four-atom intramolecular terms that are used for final tuning of the force field. The force field has been derived by using a mixture of experimental and quantum-mechanical (QM) data. The force field form is a severe approximation with many significant real energy terms not included and some even not includable in principle. Many physical forces are orthogonal to the parameter space of the used functional form, i.e., they cannot be modeled by any combination of the parameters. For example, any inclusion of polarization effects would require conformation-dependent parameters [132]. Still, despite all the known limitations, this type of analytical potential is the best atomistic model of DNA that we presently have for MD simulations.

The original AMBER force field has been refined several times, typically as a response to some problems emerging in longer simulations. All currently used versions are based on one-dimensional re-parameterizations of the dihedral energy terms. The modifications prevent some large simulation instabilities and fine-tune some properties of the force field, but they are far from being sufficient to achieve an error-free force-field performance. In other words, GQ simulations are still based on variants of a force field which was published in 1995. This is a credit to its authors but also illustrates some stagnation in force-field developments. The original 1995 force field has been designated as ff94. The ff98/99 version has improved the pucker and $\chi$ dihedral terms [133, 134], however, none of the ff94-ff99 versions proved stable in long DNA simulations. Although the ff94-ff99 versions are no longer recommended, results of some older short simulations remain valid. In particular DNA simulations have been then radically stabilized by the bsc0 modification of the $\alpha/\gamma$ dihedrals [135]. Similar major stabilization of RNA simulations has been achieved by RNA-specific $\chi_{OL3}$ modification of the $\chi$ dihedral [136]. DNA-specific $\chi_{OL4}$ parameters have improved the balance and shape of *syn-anti* DNA regions, improving GQ simulations [137]. Subsequently, $\varepsilon\zeta_{OL1}$ modified the $\varepsilon$ and $\zeta$ dihedrals, which substantially improved B-DNA helical twist and $B_I/B_{II}$ backbone dynamics [138]. The DNA-specific parametrization with similar performance as $\chi_{OL4}\varepsilon\zeta_{OL1}$ is bsc1, which modified pucker, $\chi$, $\zeta$, and $\varepsilon$ dihedrals [139]. The latest modifications profited from a new parameterization strategy matching benchmark QM and MM dihedral energy profiles with the inclusion of continuum solvent term [140], contrasting the practice of fitting the force fields using *in vacuo* computations. The most recent version provides reparameterization of the $\beta$ dihedral potential $\beta_{OL1}$, which is added to the ff99bsc0$\chi_{OL4}\varepsilon\zeta_{OL1}$ set of refinements and abbreviated as OL15 [141]. OL15 is a complete 1D reparameterization of all dihedral terms of the original Cornell et al. DNA force field. OL15 may be at the limits of tuning that can be achieved by (uncoupled) dihedral reparametrizations of the original ff94. OL15 and $\chi_{OL3}$ are the currently-recommended DNA and RNA versions of the AMBER force field in the AMBER code (i.e., AMBER16, http://ambermd.org/#AmberTools), following extensive testing [142]. Note nevertheless that all versions starting from the seminal bsc0 modification are considered as appropriate for GQ DNA simulations. The ff99bsc0$\chi_{OL4}\varepsilon\zeta_{OL1}$ [138], bsc1 [139] and OL15 [141] versions achieve very similar performance for B-DNA, though the improved β dihedral potential of OL15 appears to be advantageous for non-canonical DNAs [142].

There has been some terminology confusion in a part of the literature on the different variants of the AMBER NA force fields. It leads to potentially confusing NA force field names such as ff03 or ffxxSB, supplemented by citations to unrelated protein force fields (ff03 and ffxxSB are in fact protein force fields) or even by no citations. It is then unclear which NA force field has been applied. We suggest using the names of force field versions that were presented in the original works together with the original citations, since otherwise it is difficult to know which force field version has been used. A further overview of the general force field issues can be found elsewhere [50, 63, 142].

5. **GQ-specific limitations of the force field.**

There are at least three force-field problems specifically related to GQs, which need to be considered in studies of GQ folding. None of them can be resolved by parameterization of dihedral potentials. They are likely not correctable with the currently used force-field form, as the electronic structure effects causing them are difficult to mimic by any of the force-field terms.

5.1 **Overestimated inter-cation repulsion inside the G-stem.**

The force fields qualitatively describe the basic electrostatic stabilization of the GQ stems by monovalent ions, because long-range electrostatics is well captured by the force field. However, the force field is less satisfactory when investigating details of ion binding, which are considerably affected by polarization effects not able to be included in the force field. A known indicator of this is the occurrence of bifurcated G-quartets in simulations of some GQ topologies [143]. The ion binding energy to the quartets is underestimated and the ions inside the stem often appear to have radii that are too large [144]. This is because it is difficult to simultaneously fully balance all relevant free energy terms, in this particular case the ion hydration energies with ion – quartet interactions. Further, a recent benchmark QM study has revealed yet another and likely serious imbalance. When two or more ions are present inside the stem, their effective mutual repulsion is significantly overestimated which may cause a number of problems (Figure 14) [132]. Real ions electronically communicate with each other through polarization of the G-quartets in the stem. Thus, their mutual electrostatic repulsion is typically softened compared to interaction between two +1 point charges. To correctly describe the ion – ion interaction inside the stem, we would need to somehow reduce the charges of the ions. However, this would have to be done in a geometry-dependent manner. The charge distributions of the cations and also of the G-quartets would have to be recalculated on-the-fly for each snapshot during the simulations, rendering the fixed-charge force-field approximation useless. In reality, any change of the coordinates of the ions and of the quartets changes the electronic structure. This force-field deficiency is likely responsible for inability of MD simulations to keep bound ions at the stem-loop junctions of the d[G$_4$T$_4$G$_4$]$_2$ diagonal loop GQ [144]. These ions are expelled due to their excessive repulsion with the adjacent channel ions. Song et al. stabilized the junction ions using recalculation of the DNA charges specifically for the folded d[G$_4$T$_4$G$_4$]$_2$ GQ conformation; note that the adapted force field still remained pair-additive with fixed-charges and was thus specific for the geometry which has been used to re-calculate the charges [145].

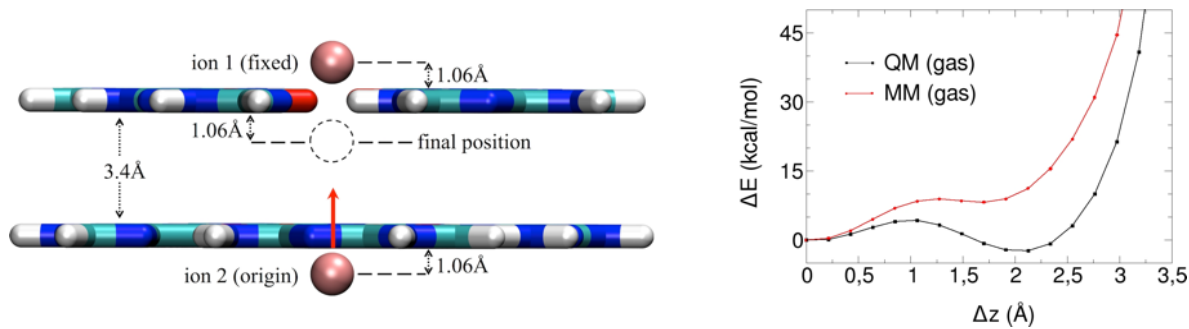

**Figure 14**. Mutual repulsion of two K$^+$ cations in the GQ channel. In the one-dimensional potential energy scan, the bottom ion moves from the origin position to the final position (left). The graph (right) shows the relative energy with respect to the starting position calculated by the AMBER (MM, red) and accurate QM calculations in the gas phase. When the cations approach each other, the force field significantly overestimates the repulsion due to lack of polarization. Adapted according to ref. [132]. Note that the purpose of this computation was not to find optimal position of the ions, but to demonstrate that pair-additive force field cannot accurately describe the energy changes associated with fluctuations and movements of the ions inside the stem.

### 5.2 Underestimated stability of GG base pairs.

Also base pairing is, in reality, associated with sizable polarization effects, manifested for example by substantial prolongations of X-H bonds participating in the pairing [146]. These electronic structure effects are not included in the force fields and thus we may expect some underestimation of base pairing energies in MD simulations. However, magnitude of the under-stabilization cannot be straightforwardly quantified, as the stability of the base pairs in MD is also affected by hydration energies. Nevertheless, we suggest that the stability of Hoogsteen base pairs in our simulations is underestimated. While this should not affect simulations of ion-stabilized GQs where the dominant stabilization comes from the ion binding, it is likely that MD simulations underestimate structural stability (i.e., life-times) of G-triplexes and G-hairpins, as we discussed above. We think that this is the resolution of the 15-TBA G-triplex issue discussed in the section 3.2, since albeit the latest dihedral force-field reparametrizations such as $\chi_{OL4}\varepsilon\zeta_{OL1}$ or OL15 (see section 4.) prolong the life-time of 15-TBA G-triplex by ~1 order of magnitude compared to bsc0, it still does not appear to be sufficient to achieve a stably folded G-triplex molecule in a converged set of simulations.

We have recently suggested that nucleic acid simulations can be selectively improved by adding small stabilizing local energy potentials to the base-pair H-bonds (Figure 15) [12]. Obviously, such a modification is structure-specific and needs to be selected for each simulated system. However, rational system-specific force field adjustments may become common in the near future, due to the above-discussed principal limitations of our capability to tune the parent multi-purpose force field. The main advantage of such structure-specific modifications is that they can be added easily and avoid undesired side-effects which often accompany attempts of general reparameterizations of the parent force field. Note that such a local potential is much milder than the commonly used restraints, targeted MD, Go-type potentials and other brute-force approaches to stabilize the desired structures. It modulates the landscape but does not eliminate the KP.

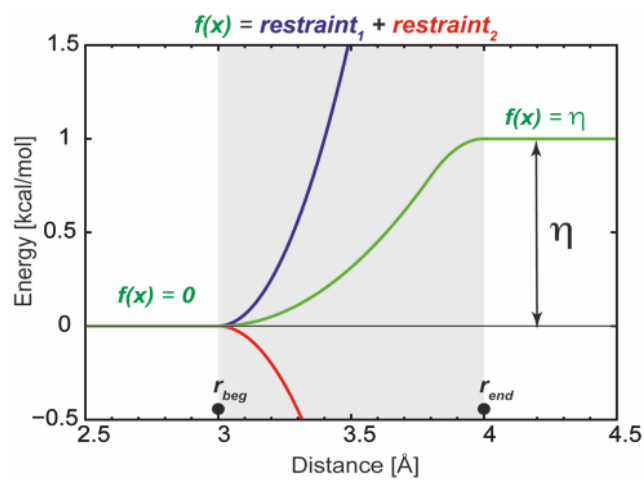

**Figure 15**. Local potential supporting strength of the hydrogen bonds (green) composed of two one-side harmonic restraints with linear extensions (blue and red curves) [12]. This local potential biases the total Hamiltonian (i.e., it creates biasing additional non-zero forces) only in narrow region of donor-acceptor distances depicted in grey. This is the region where polarization effects are primarily acting. The goal of this bias is to prolong the life-time of native H-bonds when underestimated by the parent force field. The bias does not promote folding from the unfolded states, but increases life-times of the native basins when reached spontaneously. Parameters of the bias can be adjusted, when needed, and can even be used to destabilize interactions, when reversed.

**5.3 Unstable propeller loops.**

The most perplexing GQ-specific force-field problem is the poor structural stability of the propeller loops in the simulations. Experiments suggest that propeller loops stabilize GQs and single-nucleotide loops can form only the propeller arrangement [147-152]. Parallel GQs with propeller loops are also expected to be abundant in the human genome [153]. In MD, the propeller loops remain stable in simulations of cation-stabilized GQs, except that details of their conformation are likely not fully accurate [68, 144]. This, however, is not surprising, considering the extraordinary stability of ion-stabilized GQ stems. The propeller loops (TTA as well as single-nucleotide) are, however, unstable when simulating any type of folding intermediates. In no-salt simulations of GQs and in simulation of G-triplexes, loss of the propeller loop typically initiates the unfolding [25-27]. In T-REMD simulations of G-hairpins, there is absolutely no sign of the formation, even transiently, of the parallel hairpin with a propeller loop [24]. This behavior of propeller loops is counter-intuitive considering their experimental stability and common occurrence. There are two possible explanations (or their combination) of this behavior. First, the propeller loops are formed at the very end of folding together with completion of the full ion-stabilized stem. It seems particularly difficult for parallel-stranded GQs with three such loops, though in principle they could be slowly structured from coil-like structures similar to those proposed to nucleate formation of tetramolecular GQs [79]. The second explanation is that the force field specifically underestimates the stability of the propeller loops. Although we cannot separate these two possibilities, we tend to suggest that the force field description of the propeller loops is deficient. However, we have no indications so far what energy terms could cause the force field problems, except that we know that the instability of propeller loops is insensitive to reparameterizations of the backbone dihedral potentials noted above. On the other hand, the recent electrospray mass spectrometry folding study indicates that GQ structures that contain propeller loops occur only in the late stages of folding [23]. Similarly, Raman spectroscopy suggests that a first folding product of highly concentrated $d[G_3(TTAG_3)_3]$ is an antiparallel GQ and, if unheated, it takes several days until it is converted to a parallel structure [19]. This indicates that the propeller loops may be kinetically difficult to access.

We suggest that the description of diagonal and lateral loops is considerably less problematic. Our unpublished MD simulation data (Islam et al., manuscript in preparation) indicate that the human telomeric lateral loops are well-reproduced in long simulations. They easily form in folding studies of G-hairpins [24]. Very stable simulation has been achieved for the long loop of the c-kit promoter GQ [25]. Rather notorious has been the inability of MD simulations to maintain the stability of the diagonal four-thymidine loop of the $d[G_4T_4G_4]_2$ GQ [144]. This has been linked to the loss of the bound stem-loop junction cation, which is a consequence of the non-polarizable nature of the force field (see Section 5.1). However, with the OL15 variant of the Cornell et al. force field, we have been able to stabilize the experimentally observed arrangement of this loop even despite the loss of the ions, primarily due to the improved εζ parameters [141].

6. **Concluding remarks**

We have summarized recent MD studies devoted to different aspects of the folding process of GQ molecules. When considering together with experimental data which reveal slow kinetics of the GQ folding and existence of long-living conformations, the computed results suggest that GQ folding is very different from folding of fast-folding proteins and should be described by the kinetic partitioning (KP) mechanism. Although the atomistic simulations are not robust enough to simulate the whole process, they provide important insights complementing the experiments. Thus, despite all the limitations analyzed above, we are very optimistic regarding future applications of atomistic MD to studies of various aspects of GQ folding, taking also into account recent improvements in the simulation force fields [142]. Close collaboration between experimental and theoretical groups is required to avoid over-interpretation of results. In fact, the complexity of GQ folding may turn into an advantage for the MD simulation community. GQ folding offers a much broader spectrum of interesting questions and scenarios that can be investigated, compared to the funnel-like fast-folding of small proteins or small RNAs. Our main goal is not to describe every single quantitative detail of the folding and every single micro-pathway that may exists on the folding landscape. This is still rather difficult even for fast-folding proteins, among other reasons due to challenges associated with quantitative analysis of the extreme amount of data produced by MD simulations. However, we suggest that the smart combination of ordinary simulations with various enhanced sampling methods and Markov state model approaches [48, 71, 127, 154-157] will lead to further basic insights into the GQ folding that will complement the experiments by information that is not directly accessible to experimental approaches. We also suggest that high-resolution coarse-grained simulations may provide further complementation to the atomistic MD simulations, allowing several orders of magnitude more efficient preliminary searches through the folding landscape that can then be refined by atomistic simulations [49].

Comparison of the computed results with experiments is complicated by the fact that the picture of GQ folding emerging from different experimental studies is not always mutually consistent. It is possible that in some cases the structural interpretations of the measured primary experimental data may be confusing. For example, there is at least one study directly showing that CD spectroscopy may provide false-positive results when used as a structural tool [158]. NMR measurement has revealed that $Na^+/K^+$ exchange for the d[TAG$_3$(T$_2$AG$_3$)$_3$] Htel sequence, although producing spectacular changes of CD spectra that usually are interpreted as a result of a major change in topology, are in fact not associated with any such change. There are other studies which extensively discuss why using CD spectroscopy or other low-resolution methods as a structural tool is tricky [55, 60]. CD spectroscopy is undoubtedly one of the leading tools to detect conformational behavior of GQs in solution; however, its resolution limits and the challenge of assigning topology to a particular spectrum in many instances, may occasionally bias the interpretation of the primary data. In the case noted above, without clarifying the lack of topology change by NMR measurement, CD alone would likely be interpreted as showing a major fast change of the GQ topology upon ion exchange. This would significantly obscure the folding models, as it would be in disagreement with some other experiments and with the KP model of GQ folding indirectly suggested by the computations. Some uncertainties can be expected also in structural interpretations of many other methods that are used to interrogate GQ folding. These uncertainties are assumed to typically lead to underestimation of the complexity of the GQ folding landscapes.

The MD simulations suggest that the enormous complexity of the DNA GQ folding landscape is related to the capability of many GQ-folding sequences to adopt multiple alternative GQ structures with different patterns of the *anti* and *syn* guanosines which, once formed, have long life-times. This may dominantly contribute to the partitioning of the folding landscape. When these structures are present in the course of the folding but absent in the final thermodynamic equilibrium, their detection and structural resolution are very difficult. Interestingly, in the contemporary literature, G-triplex on-pathway intermediates are probably still more popular. However, we suggest that alternative GQs (some of them not corresponding to any of the known folds) represent the dominant misfolded traps slowing down the kinetics of folding of human telomeric DNA GQs.

**Acknowledgments**

This work was supported by grant 16-13721S (JS, PS, BI) from the Czech Science Foundation. Further funding was provided by project LO1305 of the Ministry of Education, Youth and Sports of the Czech Republic (PK, PB, MO). JS acknowledges support by Praemium Academiae. SH would like to thank UCL for an Excellence Fellowship and the UCL School of Pharmacy for start-up funding. GB received funding from the European Research Council under the European Union's Seventh Framework Programme (FP/2007-2013) / ERC Grant Agreement n. 306662, S-RNA-S.